\documentclass[conference]{IEEEtran}
\IEEEoverridecommandlockouts
\usepackage{cite}
\usepackage{amsmath,amssymb,amsfonts}
\usepackage{algorithmic}
\usepackage{graphicx}
\usepackage{textcomp}
\usepackage{xcolor}
\usepackage{url}
\usepackage[bottom]{footmisc}
\def\BibTeX{{\rm B\kern-.05em{\sc i\kern-.025em b}\kern-.08em
    T\kern-.1667em\lower.7ex\hbox{E}\kern-.125emX}}
\begin{document}

\title{Deepfake audio as a data augmentation technique for training automatic speech to text transcription models}

\author{
    \IEEEauthorblockN{Alexandre R. Ferreira}
    \IEEEauthorblockA{\textit{Systems and Computing Department} \\
    \textit{Federal University of Campina Grande (UFCG)}\\
    Campina Grande, Brazil \\
    alexandre.ferreira@ccc.ufcg.edu.br}
    \and
    \IEEEauthorblockN{Cláudio E. C. Campelo}
    \IEEEauthorblockA{\textit{Systems and Computing Department} \\
    \textit{Federal University of Campina Grande (UFCG)}\\
    Campina Grande, Brazil \\
    campelo@dsc.ufcg.edu.br}
    \and
    % \and
    % \and
    % \and
}

\maketitle

\begin{abstract}
To train transcriptor models that produce robust results, a large and diverse labeled dataset is required. Finding such data with the necessary characteristics is a challenging task, especially for languages less popular than English. Moreover, producing such data requires significant effort and often money. Therefore, a strategy to mitigate this problem is the use of data augmentation techniques. In this work, we propose a framework that approaches data augmentation based on deepfake audio. To validate the produced framework, experiments were conducted using existing deepfake and transcription models. A voice cloner and a dataset produced by Indians (in English) were selected, ensuring the presence of a single accent in the dataset. Subsequently, the augmented data was used to train speech to text models in various scenarios.
\end{abstract}

\begin{IEEEkeywords}
data augmentation, deepfake audio, voice cloning, transcription models
\end{IEEEkeywords}

\section{Introduction}

Artificial intelligence has experienced significant growth in recent years due to increased computational power and the expansion of variety and volume of data exchanged over the internet. The pursuit of machine learning-generated models has expanded through various applications worldwide, such as speech-to-text transcription models. These models are utilized, for example, in translators, virtual assistants, voice search, and audio sentiment analysis \cite{speech_recognition_applications}.

For training such transcription models, labeled data is necessary, which consist of audio samples and their respective transcriptions. These transcriptions should be performed by humans to avoid biasing the results caused by transcriptions generated by another model. 

Robust transcription models should be able to generate consistent outcomes regardless of variations in a particular language (e.g., accents). However, producing such a robust model requires additional training, along with the utilization of more diversified and abundant data. 

Acquiring datasets with these characteristics is a challenging task, especially for languages less popular than English. On the other hand, producing a large dataset with these characteristics is costly and time-consuming, requiring significant financial resources and the necessary infrastructure for production. Multiple qualified individuals must manually produce the transcriptions to ensure good quality. Furthermore, to ensure transcription quality, each audio should have its transcription generated by more than one person, enabling the selection of the transcription that best represents the audio. 

One option to mitigate this problem and reduce time and cost is to use data augmentation techniques. There are various data augmentation techniques available, although most of them only allow the generation of new data with similar characteristics. For example, adding background noise or modifying the speaker's voice pitch in the audio. These techniques are efficient to produced improved transcriptors to meet certain requirements. For example, it can produce models which presents consistent results regardless of background noise or voice tone present in the input audio.

However, these data augmentation techniques do not help produce models that maintain the quality of their transcription when other characteristics vary in the input audio, such as the speaker accents. To achieve this, the model needs to be trained with data that includes a great variety of accents among speakers.

The data augmentation technique proposed in this paper is based on deepfake audio. Deepfake audio is an area of artificial intelligence that aims to produce audios that simulate the voices of specific individuals, making them sound as if they themselves had produced the audio. There are various types of models that are supposed to achieve this objective. In this paper, a model that allows voice cloning from a few seconds of audio from the original speaker is used. As a result, the data augmentation technique benefits by generating audios from the same speaker with different speech contents while preserving the voice characteristics present in the audio, such as accent.

The objective of this work is to investigate the use of this technique in datasets used for training automatic speech-to-text transcription models, evaluating the impact it has on their effectiveness. For this purpose, a framework has been implemented to investigate this technique. The framework requires a voice cloning model and a small dataset which will be submitted to the data augmentation process.

In order to validate this produced framework, various scenarios are investigated and a small dataset is used. Next, a transcription model is trained using the produced augmented dataset, which involves fine-tuning a pre-trained model. Finally, a slice of the original data is separated to evaluate the transcription model before and after training, comparing whether the training process helped the model produce better transcriptions.

The main contributions of this paper are:
\begin{itemize}
\item Provide and implement a framework able to use deepfake audio as a data augmentation technique. The implementation is ready for use and available in the repository\footnote{\url{https://github.com/alexandrerf3/data-augmentation-deepfake-audio}} of this paper. So, it is possible to execute the produced framework using any voice cloning model by replacing the component responsible for generating new audios.
\item Evaluation of a completely different scenario for data augmentation: using deepfake audio. Regarding this, no previous work was found in the literature.
\end{itemize}

Two experiments were conducted to validate the developed framework. In the first one, the framework is executed using the voice cloner with the pre-trained models provided by the author. As a result, the generated audios are used to train the transcriptor in multiple scenarios. Finally, the results were evaluated and showed that the quality of the transcriptions declined as the Word Error Rate (\texttt{WER}) metric increased by about two percent.

In the second experiment, unlike the previous one, two out of the three models used by the voice cloner were trained in different scenarios. Then the best trained model combination was selected for audio generation and subsequent training of the transcription model in multiple scenarios, like the previous experiment. Finally, the results were evaluated and showed a decline in the quality of the transcriptions, with the Word Error Rate (\texttt{WER}) metric increasing by about six percent.

The quality of the transcriptions generated by the trained transcription models in both experiments decreased in comparison with the pre-trained model. However, this decrease is believed to be due to the low quality of the audio generated by the voice cloner. Therefore, by using the produced framework and a voice cloner capable of producing high quality audio, the result should be better.

The remainder of this paper is structured as follows. The next section presents Related Work. Then Section \ref{theoretical_foundation} provides details about the Theoretical Foundation to facilitate understanding of the research. Following this, Section \ref{methodology} discusses the procedures performed for the execution of the experiments. Then Section \ref{results_discussions} describes the experiments conducted, presents and discuss the obtained results. Finally, in Section \ref{conclusions_future_work}, concludes the paper while also highlighting potential directions for future research and exploration.

\section{Related Work}

Due to the need for large datasets, several data augmentation techniques have been developed over the years. Some techniques are used to increase the data for training speech-to-text models, creating audio from modifying existing ones \cite{specaugment, specswap, speed_pertubation} or generating audio using text-to-speech models \cite{rodolfo_master_thesis}.

The audio speed perturbation technique \cite{speed_pertubation} involves modifying the audio sampling rate through the definition of an alpha value, resulting in the generation of new audios with adjusted sampling rates. This technique's efficacy has been validated through various tests.

SpecAugment \cite{specaugment} modifies the audio spectrogram using three methods: compressing/stretching the spectrogram, masking frequency channels, and masking time steps. Combined use of these methods yields good results. Similar to SpecAugment, the technique called SpecSwap \cite{specswap} swaps frequency blocks and time blocks in the audio spectrogram. While it produces good results, a comparison with SpecAugment is lacking.

Zevallos \cite{rodolfo_master_thesis} conducted data augmentation through synthetic audio and text generation. The author used Quechua language, sequence-to-sequence text generation, and text-to-speech audio generation models. The experiments produced good results and improved the transcription quality.

This paper explores data augmentation techniques for transcription model training. A framework is developed using a voice cloner model to generate new audios while preserving original dataset characteristics, such as accent. This approach provides an advantage over simpler techniques and conventional text-to-speech models, which introduce small changes/distortions or generate standardized voices without specific characteristics of the dataset.

\section{Theoretical Foundation} \label{theoretical_foundation}

This section provides details of the theoretical foundations necessary for a complete understanding of the research. First, the operation of the voice cloner chosen to be used is explained, then the chosen transcription model is described.

\subsection{Voice Cloning} \label{voice_cloning}

The chosen voice cloner for the investigations was the Real-Time Voice Cloning, provided by Corentin Jemine on his GitHub \cite{real-time_voice_cloning} and developed during his master's thesis. This cloner was selected for use due to its ability to generate new audios from a few seconds of a reference audio, without the need for retraining the models, even if the reference audio was not used during its training.

The Real-Time Voice Cloning is an implementation of the \texttt{SV2TTS} deep learning architecture \cite{sv2tts}, which consists of three independently trained components. The first component is an encoder trained on a speaker verification task using a dataset without transcriptions. It takes a few seconds of a reference audio as input and outputs a fixed-size embedding vector. The second component is a synthesizer based on \texttt{Tacotron 2} \cite{tacotron2} and is responsible for generating a mel spectrogram based on the input embedding vector and text. The third component is a vocoder, which takes the mel spectrogram as input and generates audio output. It was implemented based on \texttt{WaveRNN} \cite{wavernn} to enable real-time operation.

\begin{figure}[t]
  \centering
  \includegraphics[width=\linewidth]{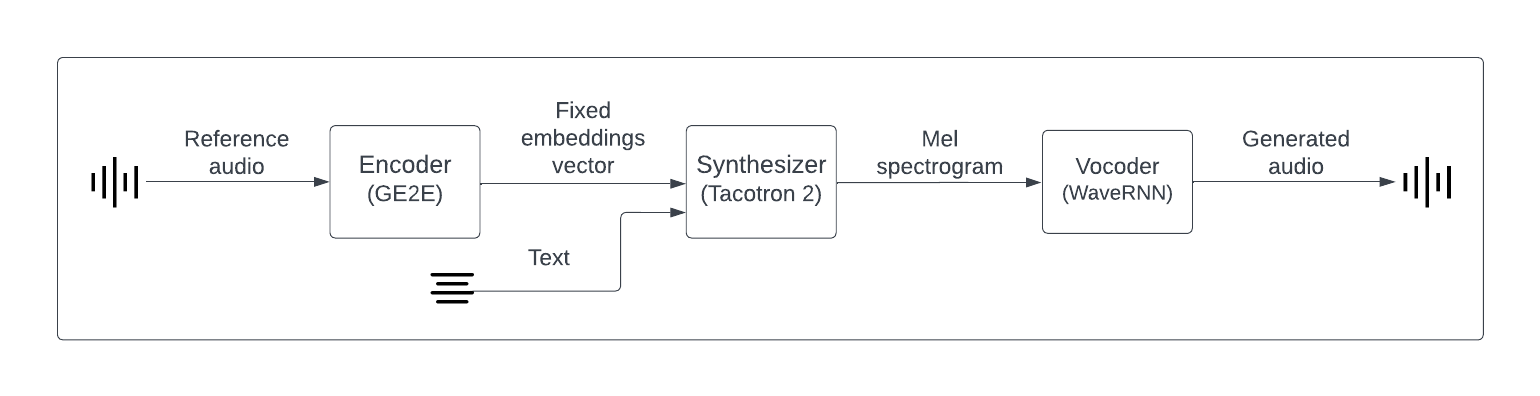}
  \caption{Voice Cloner Architecture (Real-Time Voice Cloning)}
  \label{fig:real-time_voice_cloning}
\end{figure}

Figure \ref{fig:real-time_voice_cloning} illustrates the three components with their respective inputs and outputs. In the first component, a digital representation of the voice is created, and then in the second and third components, this representation is used as a reference for generating speech from arbitrary text.

\subsection{Transcriptor}

The speech-to-text transcriptor chosen to be used in this work was DeepSpeech \cite{deepspeech}, which is an open-source speech-to-text model. The architecture of this model consists of a large Recurrent Neural Network (\texttt{RNN}). This model is simple but quite robust to background noise, speaker variation, and reverberation.

The DeepSpeech project provides pre-trained models for inference or training through transfer learning in each version.

\section{Methodology} \label{methodology}

This work consists of a qualitative experimental study. The following sections detail the procedures carried out for conducting the experiments and analyzing the results.

\subsection{Dataset}

In order to conduct the investigations, it is necessary to have a dataset that includes pairs of audio recordings with their respective transcriptions. Additionally, these recordings should be in English and spoken by individuals with the same accent. Typically, accents can vary across different regions even within the same language. Therefore, for the experiment execution, datasets recorded in English by Indian speakers were sought, as they have a distinct accent compared to American and British speakers \cite{indian_accent}.

The chosen dataset for the experiments is the NPTEL \cite{nptel} (NPTEL2020 – Indian English Speech Dataset), which was collected from YouTube videos. All the videos are in English and produced by Indians, most of them educational and with a South Asian accent. The audio from each video was extracted along with its transcription, as all the collected videos had transcriptions available that were manually uploaded by the author.

The complete \texttt{NPTEL} dataset consists of 6.2 million audio segments, with an average duration of each segment ranging from 3 to 10 seconds. It is structured in the format of LibriSpeech \cite{librispeech}, where the audio files are in \texttt{WAV} format, the transcriptions are in text files, and the metadata is in \texttt{JSON} format.

Since the \texttt{NPTEL} dataset is not manually annotated by the authors, it is not certain whether the transcriptions for each video are done manually or with the assistance of a transcription model. To address this issue, the \texttt{NPTEL} authors decided to create a sample of one thousand audios, where all of them are manually transcribed by the authors themselves. This sample is called the Pure-Set. For this reason, this portion of the data was chosen to be used as the dataset for conducting the experiments. Table \ref{tab:pure_set} show some important metadata regarding this dataset.

\begin{table}[b]
    \caption{Information about the Pure-Set dataset}
    \label{tab:pure_set}
    \centering
    \begin{tabular}{lcr}
        \hline
        Metadata & Details\\
        \hline
        Number of segments & 1000\\
        Average duration of segments & 7.82 seconds\\
        Total minutes & 130 min\\
        Dataset size & 272 MB\\
        \hline
    \end{tabular}
\end{table}

\subsection{Data Preprocessing}

\subsubsection*{Dataset Preprocessing} \label{dataset_preprocessing}

To preprocess the dataset, a script\footnote{\url{https://github.com/alexandrerf3/data-augmentation-deepfake-audio/blob/main/preprocess_nptel-pure.py}} was created to generate unique and sequential \texttt{IDs} for each file, ensuring consistency across the audios, transcriptions, and metadata. Additionally, the script utilizes the \texttt{ffmpeg-normalize} \cite{ffmpeg_normalize} library to normalize the audios, set them to a frequency of 16000 Hz, and perform post-processing steps such as noise removal and the use of a high-pass filter. Finally, the audios with empty transcriptions are removed from the dataset, and the script provides a report indicating which files were removed upon completion.

With this script, it is also possible to create subsets from the dataset by specifying the number of subsets and the number of audios in each subset. The audios for each subset are randomly selected without repetition. At the end, all the audios from the dataset are separated into the desired subsets, and text files are generated containing the \texttt{IDs} of the audios for each subset. If any audio is removed during the process due to having an empty transcription, the last subset will have a smaller number of audios.

\subsubsection*{Data Preprocessing for Cloner Training} \label{preprocessing_cloner_training}

To train the synthesizer or vocoder models of the voice cloner, additional preprocessing of the data is required. For this purpose, a script\footnote{\url{https://github.com/alexandrerf3/data-augmentation-deepfake-audio/blob/main/dataset_from_ids.py}} was created to organize the audios that will be used and place them in the file structure expected by the cloner's training scripts. It takes as input a text file containing the \texttt{IDs} of the audios and copies them, building the structure expected by the cloner.

\subsubsection*{Data Preprocessing for use in the Transcriptor} \label{preprocessing_transcriber}

Two scripts were created to preprocess the data used in the inference and training of the transcriptor. The first script\footnote{\url{https://github.com/alexandrerf3/data-augmentation-deepfake-audio/blob/main/train-deepspeech/generate_csv_files.py}} is responsible for generating \texttt{CSV} files in the format expected by DeepSpeech for training purposes. It takes a folder of audios and the number of audios to be separated for validation as input, processes them, and produces the training and validation \texttt{CSV} files. It is expected that the input audio folder contains audios generated by the voice cloner. Therefore, these audios are analyzed and compared with the original audios of their respective transcriptions during the execution of this first script.

This comparison is done to discard audios generated with poor quality because, during manual analyses, it was observed that long-duration audios generated by the voice cloner tend to have poor quality compared to the original audios of their transcriptions. The generated audios have short pauses during speech, while the original audios have longer pauses. Additionally, when the voice cloner fails to generate a particular word in an audio, it intermittently tries to generate it, producing noise until reaching the maximum duration set. Therefore, if a generated audio has a longer duration than the original audio, it can be seen as an indication that it was not generated correctly. Two attributes were defined to perform this comparison.

The first attribute is called \texttt{gap\underline{ }size\underline{ }percentage} and represents the percentage of additional duration that the generated audio must have compared to the original audio in order to be discarded. For example, using a value of 50\% for this attribute and considering that the original audio is five seconds long, the generated audio needs to have a duration of 7.5 seconds or more to be discarded. However, during some tests using this attribute, it was noticed that when the original audios were short and the generated audios were slightly longer than them, they were being discarded when they shouldn't be. For instance, considering a value of 50\% for the attribute and an original audio with a duration of two seconds, generated audios with durations of three seconds or more, based on the transcription of that original audio, were being discarded. However, after analysis, it was realized that a difference of just one second between the generated audios and the original audio was discarding audios that didn't have poor quality.

In order to mitigate this issue, a second attribute was added to be used during the comparison, called \texttt{gap\underline{ }size}. It indicates the duration by which the generated audio needs to exceed the original audio in order to be discarded. For example, considering that the original audio is seven seconds long and using a value of five for the attribute, the generated audio needs to be 12 seconds long or longer to be discarded.

The generated audio is only discarded when it exceeds the duration of the original audio from its transcription, considering both attributes in the comparison. Therefore, the discarded audios are highly likely to be generated audios with poor quality. Finally, a text file is generated with information about the discarded audios, displaying the attribute values used in the comparison, the discarded audios with their respective durations, the durations of the original audios for each transcription, and, at the end, the total number of discarded audios.

After discarding, the audios that will be part of the validation set are randomly selected without repetition, and the remaining audios are assigned to the training set. Subsequently, each transcription undergoes preprocessing, converting the text to lowercase and converting numbers into words, for example, the number 1 is transformed into 'one'. Finally, each validation and training file is created in \texttt{CSV} format following the DeepSpeech's expected model, including the respective audios and transcriptions.

The other script\footnote{\url{https://github.com/alexandrerf3/data-augmentation-deepfake-audio/blob/main/train-deepspeech/create_csv_file.py}} operates similarly to the one described earlier. It is responsible for generating \texttt{CSV} files in the format expected by DeepSpeech for audios that have not been generated by the voice cloner. For example, it can be used to generate the desired test file with audios and transcriptions that will be used to test the transcriptor. This script performs the same transcription preprocessing as the previous one and creates the \texttt{CSV} file in the desired format.

\subsection{Voice Cloner Training} \label{voice_cloner_training}

After the preprocessing performed by the script detailed in section \ref{preprocessing_cloner_training}, it is possible to use the preprocessed data for training the models used by the voice cloner. However, the chosen dataset for this work only includes audios and their respective transcriptions. Therefore, as explained in section \ref{voice_cloning}, it is only possible to train the synthesizer and vocoder models of the voice cloner.

To train these models, the scripts available in the Real-Time Voice Cloning repository \cite{real-time_voice_cloning} are used. Additionally, a step-by-step\footnote{\url{https://github.com/CorentinJ/Real-Time-Voice-Cloning/wiki/Training}} guide is provided in the same repository, which instructs on what scripts to use and in what order. For training the synthesizer model, a data preprocessing is performed using the scripts with the prefix \texttt{synthesizer\underline{ }preprocess}. Finally, the \texttt{synthesizer\underline{ }train.py} script is used for the training itself. Furthermore, for training the vocoder model, the same preprocessing steps as the synthesizer model are applied, followed by a specific vocoder preprocessing using the \texttt{vocoder\underline{ }preprocess.py} script. Finally, the training is carried out using the \texttt{vocoder\underline{ }train.py} script.

\subsection{Audios Generation} \label{audios_generation}

For generating audios using the voice cloner, two scripts were created: a main script\footnote{\url{https://github.com/alexandrerf3/data-augmentation-deepfake-audio/blob/main/generate_audios.py}} and an auxiliary script\footnote{\url{https://github.com/alexandrerf3/data-augmentation-deepfake-audio/blob/main/voice_cloning_inferences.py}}. The main script takes as input a text file containing the \texttt{IDs} of the audios used as reference audios, as well as the maximum number of audios to be generated from each reference audio. Then, for each reference audio, a random selection is made of the maximum number of other reference audios whose transcriptions will be used in the generation of the new audios.

\begin{figure}[t]
  \centering
  \includegraphics[width=\linewidth]{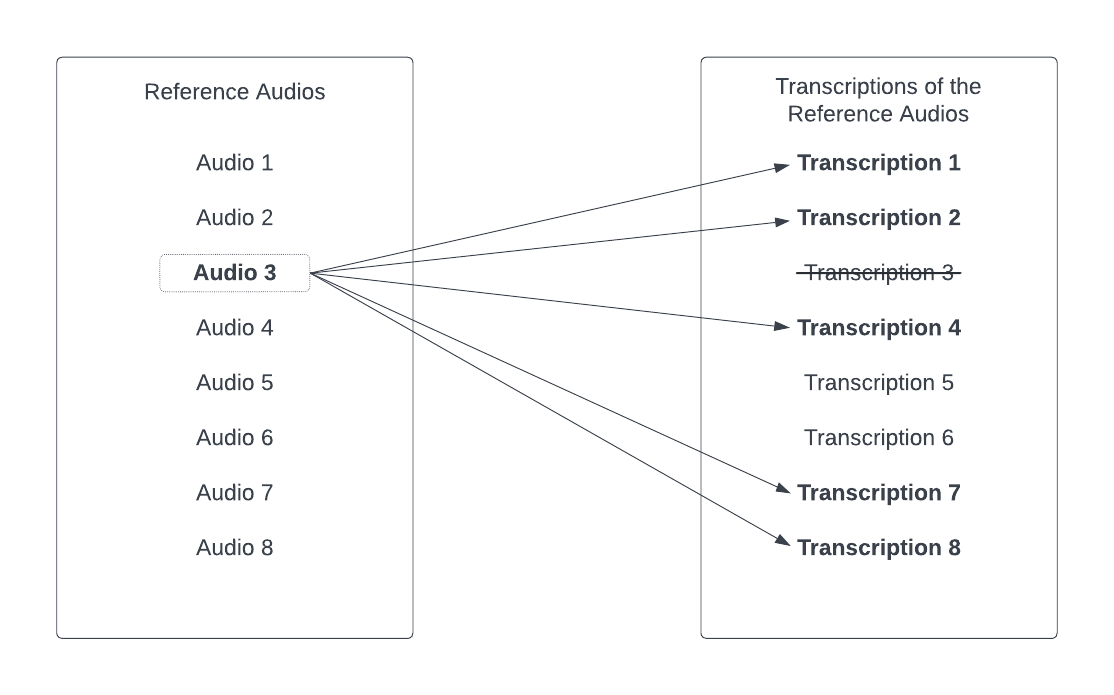}
  \caption{Illustration of a step in the process of generating new audios}
  \label{fig:audio_generation}
\end{figure}

For example, considering that the main script receives eight reference audios and a maximum limit of five, five new audios are generated for each of the eight reference audios. The text of these new audios consists of randomly selected transcriptions, without repetition, from the other reference audios, excluding the current one. Figure \ref{fig:audio_generation} illustrates a step in this example, where audio 3 is the reference audio and the highlighted transcriptions are the ones that were randomly selected for generating new audios, using the voice from audio 3 as the cloning reference. Therefore, the maximum limit of audios generated from each reference audio should be smaller than the total number of audios, considering that the transcription of the reference audio itself is not used in generating the new audios.

With the reference audios and their respective transcriptions that will be used in generating the new audios, the auxiliary script is used for applying the voice cloner. It was created based on the script called \texttt{demo\underline{ }cli.py}\footnote{\url{https://github.com/CorentinJ/Real-Time-Voice-Cloning/blob/master/demo_cli.py}} from the Real-Time Voice Cloning repository \cite{real-time_voice_cloning}. With this script, it is possible to perform inferences on the three models of the cloner in the correct order, allowing voice cloning. During the process, some audios that would have been generated may be discarded if there is an error or if the synthesizer model has generated a very small mel spectrogram.

\subsection{Training the Transcriptor}

For training the DeepSpeech transcription model, it is necessary to preprocess the data as described in section \ref{preprocessing_transcriber}. After preprocessing and generating the required \texttt{CSV} files, the training is conducted using the repository\footnote{\url{https://github.com/mozilla/DeepSpeech}} and pre-trained models\footnote{\url{https://github.com/mozilla/DeepSpeech/releases/tag/v0.9.3}} provided by DeepSpeech. The training commands for the transcriptor are listed in the repository's README\footnote{\url{https://github.com/alexandrerf3/data-augmentation-deepfake-audio/blob/main/README.md}}. By default, when training is conducted over multiple epochs, DeepSpeech evaluates the validation set at the end of each epoch and calculates a loss metric, saving the model with the lowest value. Therefore, at the end of training, the model with the lowest loss in all epochs is the one that is saved.

\subsection{Inferences in the Transcriptor}

To perform inferences in the transcriptor, a script\footnote{\url{https://github.com/alexandrerf3/data-augmentation-deepfake-audio/blob/main/deepspeech/inferences_deepspeech.py}} was developed to make this process easier. This script takes as input the model, the scorer, and a \texttt{CSV} file, formatted as expected by DeepSpeech, containing the audios used for inference and their original transcriptions.

For evaluating the transcriptions produced by the transcriptor, the Word Error Rate \cite{wer} (\texttt{WER}) metric was chosen. The \texttt{WER} is frequently employed in the performance evaluation of transcription systems, considering potential instances of word omission, addition, and substitution. Therefore, after the inferences, the original transcriptions and the transcriptions generated by the transcriptor are used to calculate the average \texttt{WER} of all the inferences made.

\section{Results and Discussions} \label{results_discussions}

In this section, we present the experiments conducted, the results obtained, and the discussions regarding them. To do this, we performed the data preprocessing described in Section \ref{dataset_preprocessing} to create two experiments using the same dataset. The experiments involve audio generation, training of the transcriptor using the generated audios, and evaluation of the transcriptor before and after training. The second experiment, unlike the first, focuses on training the cloning models to improve the results.

\subsection{Experiment 1} \label{experiment_1}

\begin{figure}[b]
  \centering
  \includegraphics[width=\linewidth]{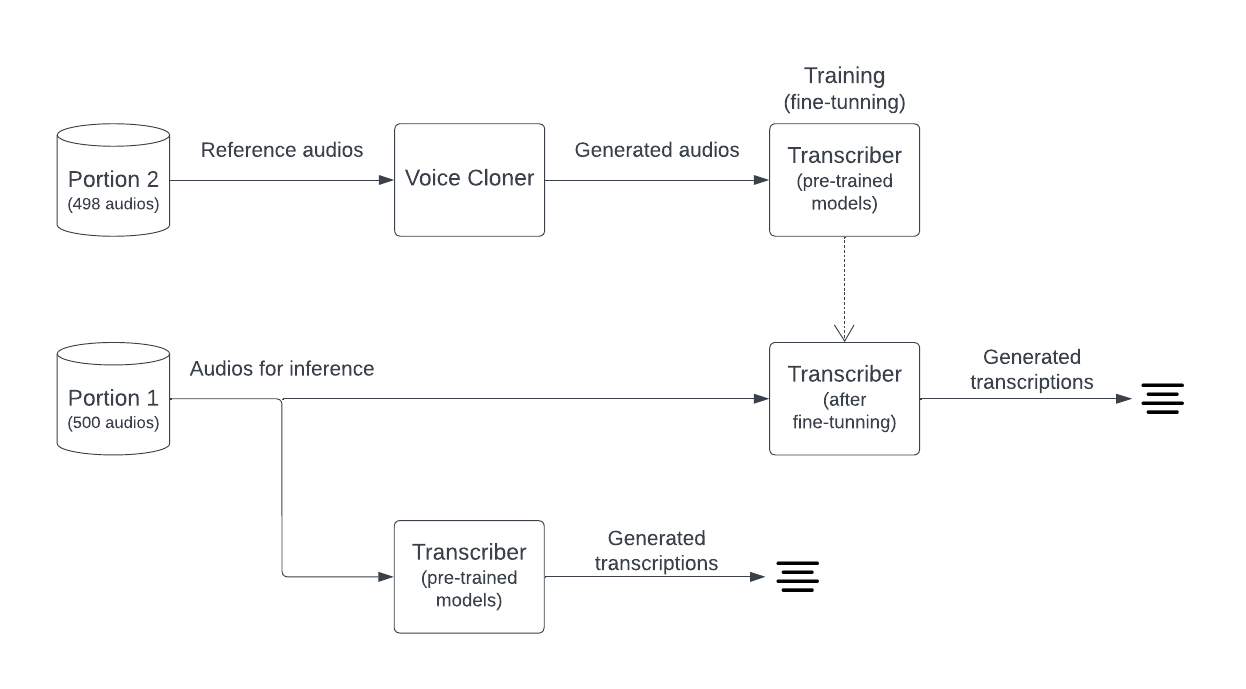}
  \caption{Illustration of the step-by-step performed in Experiment 1}
  \label{fig:experiment_1}
\end{figure}

In this experiment, the dataset is preprocessed and then split into two portions, each with 500 and 498 audios. The reduction in the second portion's size results from the discarding process during preprocessing. As a result, the first portion is used to evaluate the transcriptions generated before and after training, while the second portion is used to generate new audios used to train the transcriptor. Figure \ref{fig:experiment_1} illustrates the entire step-by-step process of this experiment.

As seen in Figure \ref{fig:experiment_1}, the second portion is used for generating new audios. In this case, the 498 audios serve as reference audios, and the limit quantity is set to 21, resulting in the generation of 10,458 audios. Subsequently, the generated audios are used to train the transcriptor in various scenarios, each consisting of 200 epochs. In each scenario, a different hyperparameter is modified to achieve better training results. Dropout is used with both the default value and a specific value of 0.4. Additionally, in one of the scenarios, the scorer is also incorporated.

Portion number 1 is used to perform inferences with the transcriptor to evaluate it. Firstly, inferences are made with the pre-trained model, and the generated transcriptions are used to calculate \texttt{WER} metric. After each model training, the portion is used again to perform new inferences and calculate a new \texttt{WER} value.

\begin{table}[b]
  \caption{Training and evaluation of transcriptor in Experiment 1}
  \label{tab:evaluation_transcriber_exp1}
  \centering
  \begin{tabular}{c c c c}
    \hline
    Scenario & Dropout & Scorer & \texttt{WER}\\
    \hline
    Pre-trained & - & - & \textbf{0.636}\\
    Fine-tuning & standard & no & 0.657\\
    Fine-tuning & 0.4 & no & 0.709\\
    Fine-tuning & standard & yes & 0.681\\
    \hline
  \end{tabular}
\end{table}

Table \ref{tab:evaluation_transcriber_exp1} displays the different training scenarios, including variations in the hyperparameters, and the corresponding \texttt{WER} results obtained for each scenario. After fine-tuning the transcription model, the \texttt{WER} result worsened compared to the pre-trained model, despite the variations in hyperparameters. After analyzing the results, it became evident that the generated audios lack satisfactory quality, with some being totally or partially incomprehensible. This factor is most likely a contributor to the observed decline in the achieved results.

\subsection{Experiment 2}

\begin{figure}[t]
  \centering
  \includegraphics[width=\linewidth]{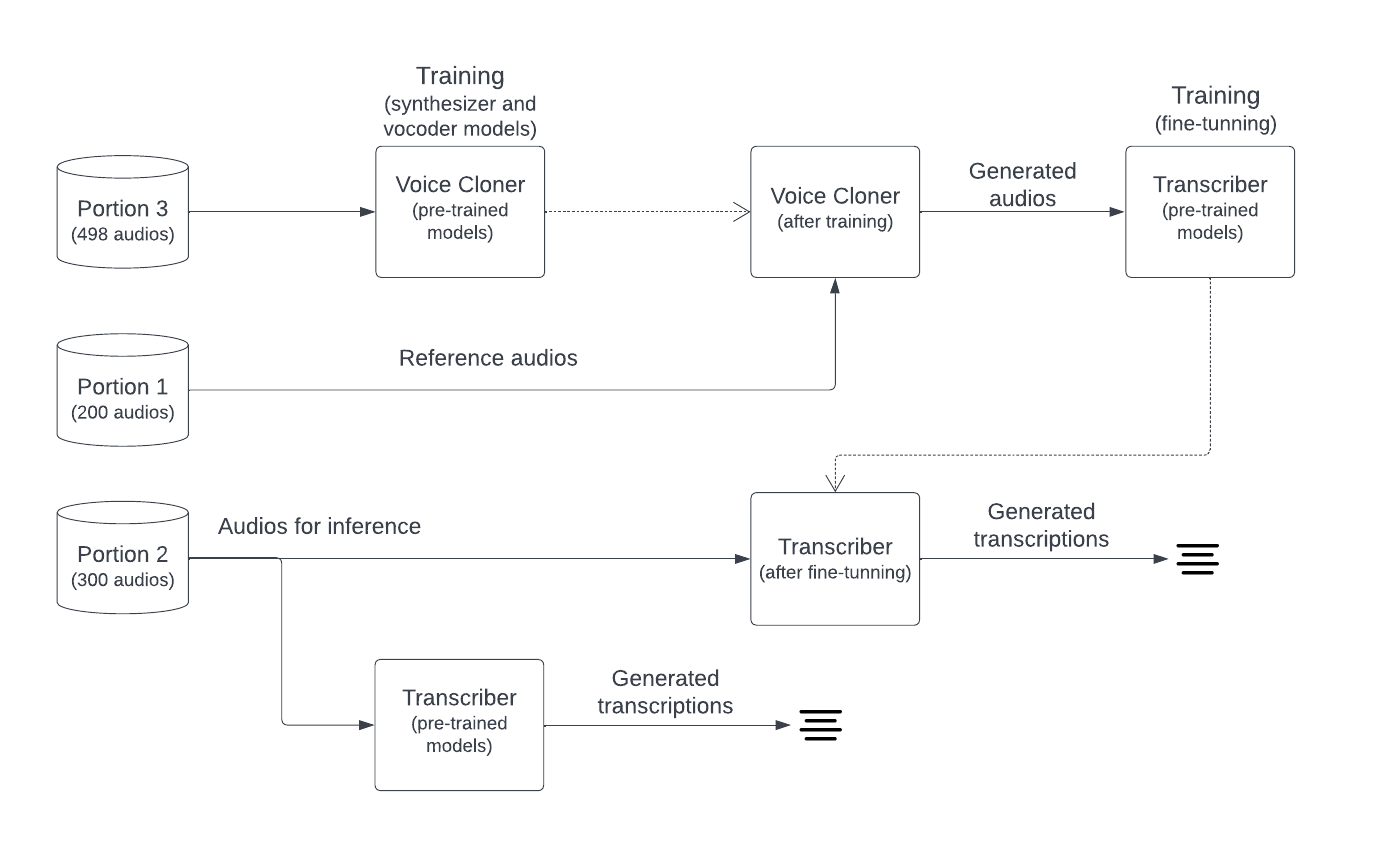}
  \caption{Illustration of the step-by-step performed in Experiment 2}
  \label{fig:experiment_2}
\end{figure}

In this experiment, the voice cloner's synthesizer and vocoder models are trained to improve the quality of the audios generated. To do this, the dataset is preprocessed and subsequently partitioned into three portions, each with 200, 300, and 498 audios, respectively. Notably, the third portion contains two fewer audios due to data discarded during the preprocessing phase. Therefore, portion number 1 is used for generating new audios, portion number 2 for evaluating the transcriptor before and after training, and portion number 3 is utilized for training the voice cloner models, as illustrated in Figure \ref{fig:experiment_2}.

To perform the training of the voice cloner models, an additional preprocessing step is required, as explained in section \ref{voice_cloner_training}. During this preprocessing, four audios were discarded from the 498 audios in portion number 3, leaving 494 audios to be used for training.

Several trainings were conducted on the synthesizer and vocoder models of the voice cloner, using combinations of fine-tuning and retraining. Table \ref{tab:training_models_voice_cloner} provides details about these trainings, showing the pre-trained models provided by the author as the default and the combinations of training performed. It also indicates the number of steps each model was trained for, highlighting the quantity of training for each combination.

\begin{table*}[h]
  \caption{Training the voice cloner synthesizer and vocoder models}
  \label{tab:training_models_voice_cloner}
  \centering
  \begin{tabular}{c c c c}
    \hline
    name & Synthesizer & Vocoder & Number of Steps\\
    \hline
    standard & pre-trained & pre-trained & 295k and 1m 159k\\
    sys\underline{ }trained & \textbf{fine-tuning} & pre-trained & \textbf{327k} and 1m 159k\\
    sys\underline{ }voc\underline{ }trained & trained (sys\underline{ }trained) & \textbf{fine-tuning} & 327k and \textbf{1m 160k}\\
    sys\underline{ }trained\underline{ }zero\underline{ }voc & trained (sys\underline{ }trained) & \textbf{retrained} & 327k and \textbf{18k}\\
    zero\underline{ }sys & \textbf{retrained} & pre-trained & \textbf{100k} and 1m 159k\\
    zero\underline{ }sys\underline{ }voc & trained (zero\underline{ }sys) & \textbf{retrained} & 100k and \textbf{13k}\\
    sys\underline{ }zero\underline{ }voc & pre-trained & \textbf{retrained} & 295k and \textbf{48k}\\
    \hline
  \end{tabular}
\end{table*}

After training the models in various combinations, it was necessary to assess the quality of the generated audios for each combination. For this purpose, a qualitative analysis is conducted. A sample of 10 audios is selected, where one audio is used as a reference, and nine audios are generated using it as a reference, resulting in a total of 90 audios. The quality of the audios is evaluated manually and classified into three categories: \textbf{poor}, \textbf{reasonable}, and \textbf{good}. Additionally, a score is calculated for each model combination based on the received classifications, where \textbf{poor} corresponds to one point, \textbf{reasonable} corresponds to two points, and \textbf{good} corresponds to three points.

\begin{figure}[h]
  \centering
  \includegraphics[width=\linewidth]{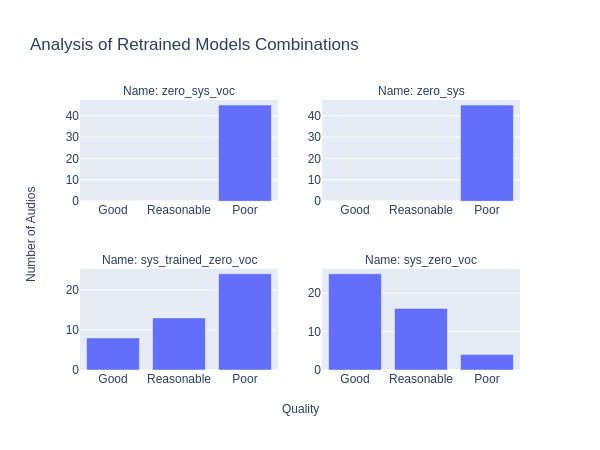}
  \caption{Qualitative analysis of the retrained models}
  \label{fig:analysis_retrained_models}
\end{figure}

In principle, these analyses are conducted on the training combinations where at least one of the models is retrained. As observed in the visualizations of Figure \ref{fig:analysis_retrained_models} and Table \ref{tab:analysis_retrained_models}, the model combinations that were retrained and achieved better results are referred to as \texttt{sys\underline{ }zero\underline{ }voc} and \texttt{sys\underline{ }trained\underline{ }zero\underline{ }voc}, while the results of the other combinations are extremely poor.

\begin{table}[b]
    \caption{Scores from the qualitative analysis of the retrained models}
    \label{tab:analysis_retrained_models}
    \centering
    \begin{tabular}{lcr}
        \hline
        Name & Score\\
        \hline
        zero\underline{ }sys\underline{ }voc & 90 points\\
        zero\underline{ }sys & 90 points\\
        \textbf{sys\underline{ }trained\underline{ }zero\underline{ }voc} & \textbf{151 points}\\
        \textbf{sys\underline{ }zero\underline{ }voc} & \textbf{202 points}\\
        \hline
    \end{tabular}
\end{table}

In the next analysis, the combinations that achieved better results in the previous analysis are considered together with the other combinations that did not have their models retrained. Furthermore, during this first analysis, it was observed that the generated audios with a long duration tend to have poor quality. Therefore, the next analysis classifies the audio duration as standard or long, aiming to verify if audios with a long duration indeed tend to be poor.

\begin{figure}[h]
  \centering
  \includegraphics[width=\linewidth]{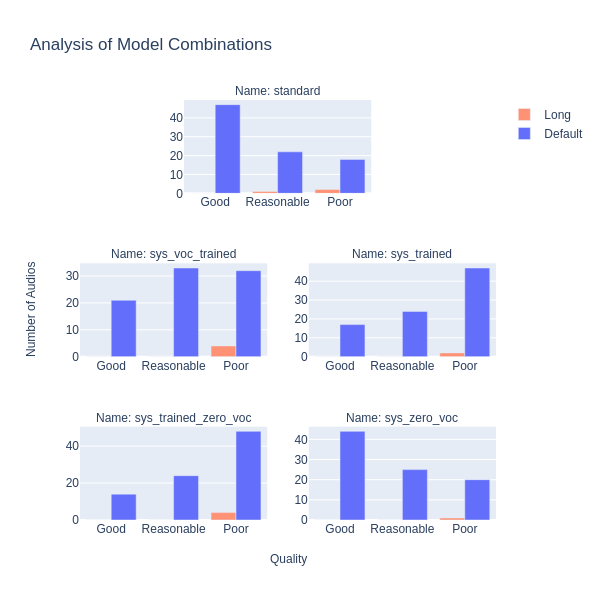}
  \caption{Qualitative analysis of the models}
  \label{fig:analysis_models}
\end{figure}

\begin{table}[b]
    \caption{Scores from the qualitative analysis of the models}
    \label{tab:analysis_models}
    \centering
    \begin{tabular}{lcr}
        \hline
        Name & Score\\
        \hline
        \textbf{standard} & \textbf{207 points}\\
        sys\underline{ }voc\underline{ }trained & 165 points\\
        sys\underline{ }trained & 148 points\\
        sys\underline{ }trained\underline{ }zero\underline{ }voc & 142 points\\
        \textbf{sys\underline{ }zero\underline{ }voc} & \textbf{203 points}\\
        \hline
    \end{tabular}
\end{table}

After the last qualitative analysis of the models, it can be observed in the visualizations of Figure \ref{fig:analysis_models} and Table \ref{tab:analysis_models} that the combinations of models that achieved better results are the ones called \texttt{standard} and \texttt{sys\underline{ }zero\underline{ }voc}. The combination of models called \texttt{standard} was already used in Experiment 1 (\ref{experiment_1}) for audio generation, transcriptor training, and analysis of the results. Therefore, in this experiment, the combination of models called \texttt{sys\underline{ }zero\underline{ }voc} is used for further investigations.

In addition, this last analysis allowed us to verify if long-duration audios tend to have poor quality. Thus, observing Figure \ref{fig:analysis_models}, it can be affirmed that the majority of audios with long duration do indeed have poor quality. Therefore, the discarding of long-duration audios is valid, and it is done during the preprocessing of the generated audios, before they are used in the transcription model, as described in Section \ref{dataset_preprocessing}.

The models from the combination named \texttt{sys\underline{ }zero\underline{ }voc} are then used in the voice cloner to generate new audios, based on the 200 reference audios from portion number 2 and with a limit quantity set to 52, resulting in a total of 10,400 audios. The limit quantity value of 52 is chosen aiming to generate an approximate number of audios similar to what was generated and used in experiment 1.

The generated audios are then used in the training of the transcription model in different scenarios, varying its hyperparameters as conducted in experiment 1 (\ref{experiment_1}).

Portion number 3 of the data is used to evaluate the transcription model before and after the trainings, in different scenarios. The 300 audios from this portion are used to make inferences with the various models, calculating the \texttt{WER} for each one. In table \ref{tab:evaluation_transcriber_exp2}, you can observe the models, the variations of the hyperparameters, and the \texttt{WER} value for each model.

\begin{table}[b]
  \caption{Training and evaluation of transcriptor in Experiment 2}
  \label{tab:evaluation_transcriber_exp2}
  \centering
  \begin{tabular}{c c c c}
    \hline
    Scenario & Dropout & Scorer & \texttt{WER}\\
    \hline
    Pre-trained & - & - & \textbf{0.648}\\
    Fine-tuning & standard & no & 0.710\\
    Fine-tuning & 0.4 & no & 0.742\\
    Fine-tuning & standard & yes & 0.711\\
    \hline
  \end{tabular}
\end{table}

After fine-tuning the transcription model using the audios generated by the voice cloner with the new models, the transcriptions significantly worsened, and the \texttt{WER} metric increased by approximately six percent. A probable cause for the decline in results is the quality of the generated audios, which, even after several attempts to train the voice cloner models, continue to have poor quality.

One option to improve the results is to change the voice cloner. However, for the investigations and experiments conducted in this work, a voice cloner capable of cloning a voice from a few seconds of a reference audio is required. As a result, Real-Time Voice Cloning \cite{real-time_voice_cloning} was the only option found with freely available code for use. Other options that claimed to have higher quality in voice cloning do not have their codes available due to the potential misuse of such technology. Additionally, some sources mention that the codes will only be disclosed once reliable detectors for audio generated from deepfake techniques are developed.

Furthermore, the authors of the \texttt{SV2TTS} architecture \cite{sv2tts}, used in Real-Time Voice Cloning, point out that the most efficient and effective way to improve the quality of generated audios is to train the encoder model, as can be observed in the cloner's architecture in Figure \ref{fig:real-time_voice_cloning}. However, during the course of this work, it was not possible to train the encoder model because it requires a dataset where speaker information is available for each audio. Unfortunately, the dataset used in this work does not provide such information.

Another factor that possibly influences the lack of improvement in the cloner after the training is that the audios in the dataset used in this work are noisy. They are extracted from YouTube\footnote{YouTube — www.youtube.com} videos recorded in various environments with different recording equipment. Furthermore, since most of the videos are educational, a significant portion of the speech in the audios contains technical language related to the taught content. Some examples of transcriptions from the audios, observed during the manual analysis, can be seen in Table \ref{tab:transcriptions}.

\begin{table}[b]
  \caption{Examples of transcriptions of the audios from the dataset used in the experiments}
  \label{tab:transcriptions}
  \begin{tabular}{c}
    \hline
    Transcription\\
    \hline
        \footnotesize
        NOW THIS PREFERENTIAL FLOW OF CURRENT IN\\
        \footnotesize
        A DIODE IS UTILISED TO CONVERT AN AC TO DC SUPPOSE\\[5pt]
        \hline
        \footnotesize
        AND WHAT IS FIRST OF F FIRST OF F IS\\
        \footnotesize
        LEFT PARENTHESIS AND IDEALS SO\\ [5pt]
        \hline
        \footnotesize
        Y 1 ONE D X IF THE DIVIDED DIFFERENCE TERM\\
        \footnotesize
        IS ZERO ERROR IS GOING TO BE EQUAL TO ZERO\\ [5pt]
    \hline
  \end{tabular}
\end{table}

These data, which contain more technical language, are not commonly found in datasets. Therefore, it is highly likely that the pre-trained models of the voice cloner and the transcription were not trained with such technical words.

\section{Conclusions and Future Work} \label{conclusions_future_work}

To conduct the investigations and experiments in this work, using deepfake audio as a data augmentation technique, we sought a dataset in the English language that exclusively featured the Indian accent. This dataset needed to contain pairs of audios with their respective transcriptions. Additionally, it was necessary to find a voice cloner capable of cloning voices from a few seconds of a reference audio. This allows for the augmentation of the chosen dataset.

With the augmented dataset in hand, it was necessary to verify whether its utilization in training a transcriptor would result in transcriptions of higher quality. For this purpose, the transcription model called DeepSpeech was employed to conduct the investigations. By selecting a subset of the data, inferences were made to the transcriptor, and the quality of its generated transcriptions was measured using a metric called \texttt{WER} (Word Error Rate). Subsequently, after training the transcription model using the augmented data, the same subset was used to make new inferences, aiming to assess the quality of the transcriptions produced after training and determine whether there was an improvement in the results or not.

With the experiments conducted in this work, no improvements were observed in the quality of the transcriptions generated after training the transcription model. Despite training the transcriptor in various scenarios, all of them showed a deterioration in transcription quality. One likely reason for these results is the quality of the audios generated by the voice cloner, as manual analyses of the audios revealed poor quality. Even after training some of the models used by the voice cloner, the quality of the generated audios remained unsatisfactory. Therefore, the audios generated with poor quality may be hindering the learning of the transcription model.

In an attempt to achieve better results, a future work that can be conducted is improving the quality of the generated audios. For this purpose, one can seek better training of the voice cloner models by making changes to the hyperparameters or architectures of the synthesizer and vocoder. Additionally, it would be beneficial to find a dataset in the English language that specifically includes Indian accents and provides speaker identification. This would allow for the training of the encoder model of the voice cloner, thereby improving its performance.

Additionally, as voice cloning models are constantly evolving, it is possible to use the framework developed throughout this work in conjunction with a new voice cloning model. This new voice cloning model should be suitable for conducting the experiments, and if it has better audio generation quality, it will likely yield better results.

The dataset explored in this work has some characteristics that make it challenging to generate audios and transcriptions, such as background noise and technical language. One possible improvement for conducting the experiments is to find or create a dataset with a larger quantity of audios, where they have less noise and a less technical language.

\bibliographystyle{ieeetr}
\bibliography{references}

\end{document}